\documentclass[aps,twocolumn,a4paper,showpacs]{revtex4}
\usepackage{graphicx}
\usepackage[all]{xy}
\usepackage{amsmath}
\usepackage{amssymb}
\usepackage{color}
\usepackage[latin1]{inputenc}

\newcommand{\bb}{\bibitem}
\newcommand{\bes}{\begin{subequations}}
\newcommand{\ees}{\end{subequations}}
\def\ben{\begin{eqnarray}}
\def\een{\end{eqnarray}}
\def\be{\begin{equation}}
\def\ee{\end{equation}}

\begin{document}
\title{Solitonic traveling waves in Galileon theory}
\author{D. Bazeia$^{1,2}$, L. Losano$^{1,2}$, and J.R.L. Santos$^2$}
\affiliation{ $^1${Departamento de F\'\i sica, Universidade Federal da Para\'\i ba, 58051-970 Jo\~ao Pessoa PB, Brazil}\\
$^2${Departamento de F\'\i sica, Universidade Federal de Campina
Grande, 58109-970 Campina Grande PB, Brazil} }
\date{\today}
\begin{abstract}
This work deals with traveling waves in the two-dimensional Galileon theory. We use the Hirota procedure to calculate one-Galileon, two-Galileon, three-Galileon and breather-like Galileon solutions in the theory under consideration. The results offer strong evidence that the Galileon traveling waves are solitons, and that the Galileon theory is integrable.
\end{abstract}
\pacs{02.30.Jr; 05.45.Yv}

\maketitle

{\it Introduction.} In this work we deal with relativistic models described by a single real scalar field in two-dimensional space-time. The study is inspired on the galileon field, that is, a real scalar field that engenders Galileo invariance, such that, if $\phi=\phi(x)$ is real, it is a galileon field if its Lagrange density is symmetric under the Galileo transformation $\phi\to\phi+a+ b{\cdot} x$, with $a$ being a constant scalar and $b$ a constant vector. 

The galileon field was studied in \cite{g1,g2} aimed to investigate self-accelerating solutions in the absence of ghosts, and has been further investigated in a diversity of contexts, with direct phenomenological applications, as one can see in the recent reviews \cite{r1,r2,r3}. In particular, in \cite{gs1,gs2} the authors study the presence of solitons: in \cite{gs1} it is shown that the galileon field cannot give rise to static solitonic solutions; however, in \cite{gs2} the authors study the presence of soliton-like traveling waves for the galileon field in two-dimensional space-time. 

The study \cite{gs2} shows the presence of traveling wave analytically, and the numerical investigation suggests that the traveling wave behaves as solitons. This investigation has motivated us to further study the problem, and here we give another, stronger evidence that the traveling wave is a soliton. We do this working with the Hirota procedure \cite{H1,H2,HI,HB,H,hz}, and we also construct new solutions, of the multi-soliton type. 

To be specific, we focus on the model
\be \label{eq1}
{\cal L}= \partial_\mu\phi\partial^\mu\phi + \alpha \,\partial_\mu\phi\partial^\mu\phi\; \Box\phi,
\ee
where $\alpha$ is a positive real parameter. We are using $\Box \equiv g^{\mu\nu}\partial_\mu\partial_\nu \phi $, the metric is diagonal
$(+,-)$ and the scalar field, space and time coordinates, and the coupling constant $\alpha$ are all dimensionless.

This model has the equation of motion
\be \label{eq2}
\partial_{\mu}\partial^{\mu}\phi+\alpha\,\left[(\Box\phi)^{2}-\partial_{\mu}\partial_{\nu}\,\phi\,\partial^{\mu}\partial^{\nu}\,\phi\right]=0\,,
\ee
and for $\phi=\phi(x,t)$, we can write, explictly,
\be \label{eq3}
\phi_{t\,t}-\phi_{x\,x}+2\,\alpha\,\left[\phi_{t\,x}^{\,2}-\phi_{t\,t}\,\phi_{x\,x}\right]=0\,.
\ee
We suppose that $\phi(x,t)$ is a traveling wave, in the form $\phi(x,t)=\phi(kx-wt)$. In this case, we get
\be \label{eq4}
\phi_t=-\frac{\omega}{k}\,\phi_x\,;\;\;\;  \phi_{t\,t}=\frac{\omega^2}{k^2}\,\phi_{x\,x}\,; \;\;\; \phi_{t\,x}=-\frac{\omega}{k}\,\phi_{x\,x}\,,
\ee
and the equation of motion leads to
\be \label{eq5}
\left(\frac{\omega^2}{k^2}-1\right)\,\phi_{x\,x}=0\,,
\ee
such that $k=\pm\,\omega$. Thus, any well-behaved traveling wave of the form $\phi_\pm(x,t)=\phi(x\pm t)$ solves the equation of motion and is a Galileon solution; see, e.g., Ref.~\cite{gs2}.

Here we further study the Galileon theory, but we follow an alternative route, using the Hirota bilinear method \cite{H1}, from which we obtain new solutions. We have chosen this route because the Hirota method has been very effective in the construction of soliton solutions to several integrable models. As one knows, it is possible to construct one- and two-soliton solutions even for non-integrable models, but the
existence of three-soliton solution strongly suggests equivalence to integrability \cite{H,hz}. 

We organize the work as follows. We start briefly reviewing the Hirota method for the sine-Gordon equation in the next Section, where we construct one- and two-soliton solutions. We then study the Galileon model, and construct one-, two- and three-Galileon solutions, and another solution, of the breather type. We end the work with some comments on the main results.

{\it Hirota for sine-Gordon.} Let us briefly review the application of the Hirota bilinear method when we consider the sine-Gordon equation.
The sine-Gordon model is based on the second order partial differential equation
\be \label{eq6}
\phi_{\,x\,x}-\phi_{\,t\,t}=\sin\,\phi\,.
\ee
In order to apply the Hirota method, we introduce $f(x,t)$ and $g(x,t)$ and work with the transformation
\be \label{eq7}
\phi(x,t)=4\,\arctan\,\left(\frac{g}{f}\right)\,.
\ee
We remark that \eqref{eq7} was also used to investigate the modified Korteveg-de Vries equation \cite{HI}. Now, after substituting this transformation into the sine-Gordon equation, we obtain the pair of equations 
\ben
&& \label{eq8}
\left(D_x^{\,2}-D_t^{\,2}-1\right)\,(f\,.\,g)=0\,, \\
&& \label{eq9}
\left(D_x^{\,2}-D_t^{\,2}\right)\,(f\,.\,f-g\,.\,g)=0\,,
\een
where $D$ stands for the Hirota derivative, such that $D^n_x$ is defined as
\ben \label{eq10}
D_x^{\,n}\,(f.g)&=&\left(\frac{\partial}{\partial\,y}-\frac{\partial}{\partial\,z}\right)^{\,n}\,(f(y).g(z))\big|_{y=z=x},\nonumber\\
&=&\partial_y^nf(x+y)g(x-y)\big|_{y=0}.
\een
We note that the action of $D$ on the product is similar to the Leibniz rule, but with an important change of sign. 
The two bilinear equations \eqref{eq8} and \eqref{eq9} control the Hirota procedure for the sine-Gordon equation, and can be used to find multi-soliton solutions. To see this explicitly, we take

\be \label{eq11}
f=f_0+\epsilon\,f_1+\epsilon^2\,f_2+...\,;\,\,\, g=g_0+\epsilon\,g_1+\epsilon^2\,g_2+...\,,
\ee
and use the Eqs.~(\ref{eq8}) and (\ref{eq9}) to find the relations, to $O(\epsilon)$ and $O(\epsilon^2)$, respectively,
\ben \label{eq12}
&& 
\left(D_x^{\,2}-D_t^{\,2}-1\right)\,(f_1\,g_0+g_1\,f_0)=0\,; \\ 
&&
\left(D_x^{\,2}-D_t^{\,2}-1\right)\,(f_2\,g_0+f_1\,.\,g_1+f_0\,g_2)=0\,, \nonumber
\een
and
\ben \label{eq13}
&&
\left(D_x^{\,2}-D_t^{\,2}\right)\,(f_0\,f_1-g_0\,g_1)=0\,; \\
&&
\left(D_x^{\,2}-D_t^{\,2}\right)\,(2\,f_0\,f_2+f_1\,f_1-2\,g_0\,g_2-g_1\,g_1)=0\,. \nonumber
\een

\subsection{One-Soliton Solution}

In order to derive one-solition solution from the above bilinear relations, we work with 
\be \label{eq14}
f=f_0=1\,;\qquad g=\epsilon\,g_1\,;\qquad g_1=e^{\,\theta_1}\,,
\ee
where $\theta_1=k_1\,x+\omega_1\,t$, therefore $f_i=0$ for $i>0$, $g_0=0$ and $g_i=0$ for $i>1$. After substituting these $f$ and $g$ into the Eqs. (\ref{eq12}) and (\ref{eq13}), we have $k_1^{\,2}-\omega_1^{\,2}-1=0$ or 
\be \label{eq15}
\omega_1=\pm\,\sqrt{k_1^{\,2}-1}\,,
\ee
and we obtain
\be \label{eq16}
\phi(x,t)=4\,\arctan\,\left(e^{k_1\,x+\omega_1\,t}\right)\,,
\ee
where we have taken $\epsilon=1$.

\subsection{Two-Soliton Solution}

To get to the two-soliton solution, we consider
\be \label{eq17}
f=1+\epsilon^{\,2}\,f_2\,;\qquad g=\epsilon\,g_1\,;\qquad g_1=e^{\,\theta_1}+e^{\,\theta_2}\,,
\ee
where $f_2$ is in principle unknown, $\theta_i=k_i\,x+\omega_i\,t$ for $i=1,2$. Furthermore, $f_1=0$, $f_i=0$ for $i>2$, $g_0=0$ and $g_i=0$ for $i>1$. As in the one-soliton case, we substitute the above expressions into Eqs.~(\ref{eq12}) and (\ref{eq13}) in oder to obtain the dispersion relations $\omega_i=\pm\,\sqrt{k_i^{\,2}-1}$, $i=1,2$. Moreover, the function $f_2$ obeys
\be \label{eq18}
f_{\,2\,x\,x}-f_{\,2\,t\,t}+\left[(\omega_1-\omega_2)^2-(k_1-k_2)^2\right]\,e^{\,\theta_1+\theta_2}=0\,,
\ee
which can be solved analytically, giving
\be \label{eq19}
f_2=A\,e^{\,\theta_1+\theta_2}\,, \qquad A=\frac{(k_1-k_2)^2-(\omega_1-\omega_2)^2}{(k_1+k_2)^2-(\omega_1+\omega_2)^2}\,,
\ee
where $A$ is the phase factor. Consequently, by setting $\epsilon=1$ the final form of the two-soliton solution is given by 
\be \label{eq20}
\phi=4\,\arctan\,\left(\frac{e^{\,\theta_1}+e^{\,\theta_2}}{1+A\,e^{\,\theta_1+\theta_2}}\right)\,,
\ee
as it is known.

{\it Hirota for Galileons.} Let us now concentrate on Galileons. Here we introduce the same transformation used in the sine-Gordon case, that is,
\be \label{eq21}
\phi=\arctan\,\left(\frac{g}{f}\right)\,.
\ee
We substitute \eqref{eq21} in the equation of motion for Galileons to get to the equations
\be \label{eq22}
\left(D_x^{\,2}-D_t^{\,2}\right)\,(f\,.\,g)=0\,;
\ee
\be \label{eq23}
\left(D_x^{\,2}-D_t^{\,2}\right)\,(f\,.\,f-g\,.\,g)=0\,;
\ee
\be \label{eq24}
\left(D_x^{\,2}\mp\,D_t\,D_x\right)\,(f\,.\,g)=0\,;
\ee
\be \label{eq25}
\left(D_x^{\,2}\mp\, D_t\,D_x\right)\,(f\,.\,f-g\,.\,g)=0\,.
\ee
These bilinear equations constitute the key result of the Hirota procedure for Galileons, and can be used to find multi-soliton solutions.
Before searching for solitons, however, we note that equations \eqref{eq22}-\eqref{eq25} remain invariant under the transformations ($\mu$ and $\nu$ constants)
\ben
&&f\to e^\mu  f;\;\;\;g\to e^\nu  g,\label{eq26a}
\\
&&f\to e^{(\mu x+\nu t)} f;\;\;\; g\to e^{(\mu x+\nu t)} g,\label{eq27a}
\\
&&f\to e^{(\mu\partial_x+\nu\partial_t )} f;\;\;\; g\to e^{(\mu\partial_x+\nu\partial_t)} g.\label{eq28a}
\een 
This is an important result, indicating the existence of a Lie algebra underlying the bilinear equations, strongly suggesting integrability  of the Galileon equation \cite{HB,H}.

Turning attention to the soliton solutions, to see how the bilinear equations \eqref{eq22}-\eqref{eq25} work for Galileons, we use them and the functions introduced in Eq.~(\ref{eq11}) to find analytical solutions, as we do below.

\subsection{One-Galileon Solution}
Here we work with
\be \label{eq26}
f=f_0\,;\qquad g=\epsilon\,g_1\,; \qquad g_1=e^{\theta_1}\,,
\ee   
where $\theta_1=k_1\,x+\omega_1\,t$, $f_0$ is a constant, $f_1=0$, $g_0=0$ and $f_i=g_i=0$ for $i>1$. Thus, from equation (\ref{eq22}) we find the dispersion relation $\omega_1=\pm\,k_1$.
Moreover, equations (\ref{eq23}), (\ref{eq24}) and (\ref{eq25}) are all satisfied.
Therefore, one finds one-Galileon solutions as
\be \label{eq27}
\phi_{\,\pm}=\arctan\,\left(\frac{e^{\,k_1\,x \pm k_1\,t}}{f_0}\right)\,.
\ee

\subsection{Two-Galileon solution}

\begin{figure}[h]
\includegraphics[{width=1.05\columnwidth}]{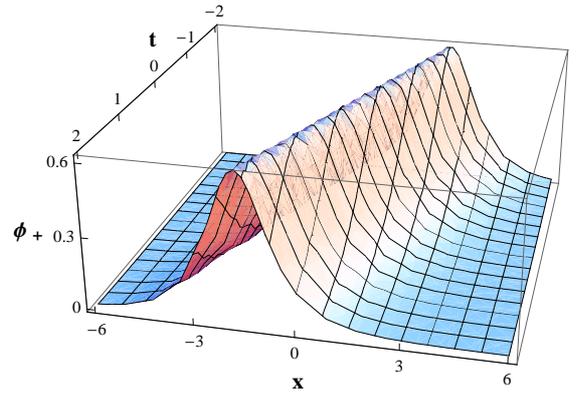}
\caption{We depict an example of the two-Galileon solutions, where we consider $\phi_{\,+}(x,t)$ of Eq.~(\ref{eq30}), with $a=1$, $b=1$, $f_0=1$, $B=2$, $k_1=-2$, $k_2=-1$.}
\label{fig1}
\end{figure}

In analogy with the sine-Gordon model, we consider
\be \label{eq29}
f=f_0+\epsilon^2\,B\,e^{\,\theta_1+\theta_2}\,;\qquad g_1=a\,e^{\theta_1}+b e^{\theta_2}\,,
\ee
where $a$ and $b$ and $B$ are arbitrary constants, $\theta_i=k_i\,x + \omega_i\,t$, $f_1=0$, $g_0=0$ and $f_i=g_i=0$ for $i>2$. We can directly prove that all the bilinear operators are satisfied by these forms of $f(x,t)$ and $g(x,t)$ with the dispersion relations $\omega_i=\pm\,k_i$, $i=1,2$. Another feature is that $B$ is now arbitrary, leading to two-Galileon solutions of the form
\be \label{eq30}
\phi_{\,\pm}=\arctan\,\left[\frac{a\,e^{\theta_1}+b\,e^{\theta_2}}{f_0+B\,e^{\theta_1+\theta_2}}\right]\,,
\ee
which we illustrate in Fig.~\ref{fig1}.

We note that the one-Galileon solutions can be recovered if we set $e^{\theta_2}\rightarrow 0$ in the above two-Galileon results; it can also be obtained in  the limit where the second wave is far away, in the limit $e^{\theta_2}\rightarrow \infty$. These behaviors corroborates with the analysis done in Ref.~\cite{hz} in the context of soliton solutions.

\begin{figure}[h]
\includegraphics[{width=1.05\columnwidth}]{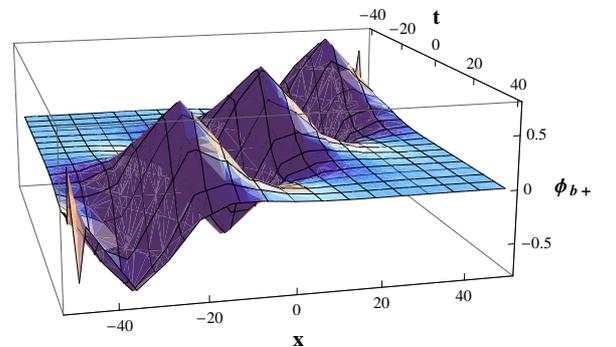}
\vskip 0.3 cm
\includegraphics[{width=0.85\columnwidth}]{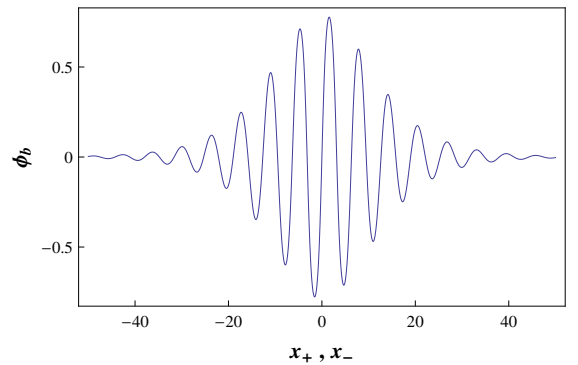}
\caption{In the top panel we depict $\phi_{b+}(x,t)$, where we choose $f_0=1$ and $\rho=0.993$. In the bottom panel we depicted $\phi_{b+}$ in terms of the light-cone coordinates  $x_{\pm}=x\,\pm\,t$.}
\label{fig2}
\end{figure}

An interesting case emerges from the two-Galileon solutions (\ref{eq30}) when we choose $a=1$, $b=-1$, $B=1$ and $k_2=1/k_1$, leading us to
\be \label{eq32}
\phi_{\,\pm}=\arctan\,\left[\frac1{f_0}\frac{\sinh\,\left(\frac{k^{2}-1}{2\,k}\,(x\pm t)\right)}{\cosh\,\left(\frac{k^{2}+1}{2\,k}\,(x\pm t)\right)}\right]\,,
\ee
where we have changed $k_1\to k$. Now, we define
\be \label{eq33}
\rho=i\,\frac{1-k^{2}}{2\,k}\,;\qquad \sqrt{1-\rho^2}=\frac{k^{2}+1}{2\,k}\,,
\ee
and we consider $\rho$ real, which allows to write the above solutions in the form, after changing $f_0\to if_0$,
\be \label{eq34}
\phi_{b\pm}=\arctan\,\left[\frac{1}{f_0}\,\frac{\sin\,\left(\rho\,(x\pm t)\right)}{\cosh\,\left(\sqrt{1-\rho^2}\,(x\pm t)\right)}\right]\,.
\ee
The procedure here is similar to the case of breather in the sine-Gordon model, so the solutions $\phi_{b\pm}$ are breather-like solutions, analogous to the breather solution of the sine-Gordon model. In Fig.~\ref{fig2} we depict an example of a breather-like Galileon.

\begin{figure}[h]
\includegraphics[{width=1.0\columnwidth}]{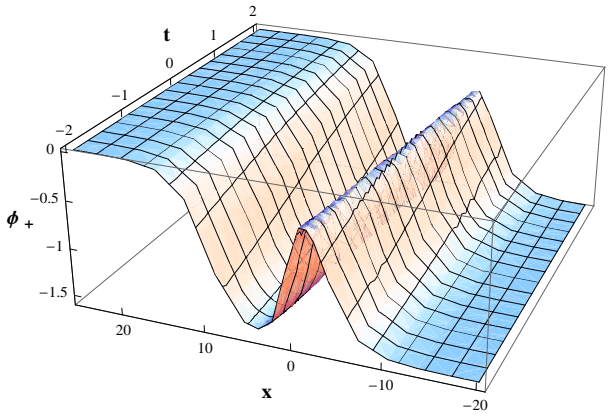}
\caption{An example of the three-Galileon solutions, where we consider $\phi_{\,+}(x,t)$ with $a=0.5$, $b=5$, $c=0.125$, $C=5$, $f_0=1$,
$A_1=-0.1$, $A_2=-2$, $A_3=-2$, $k_1=0.5$, $k_2=1$, $k_3=-1$.}
\label{fig3}
\end{figure}

\subsection{Three-Galileon solution}

The next {\it Ansatz} to determine the three-Galileon solutions is based on the same choice (\ref{eq21}), but now we take
\be \label{eq35}
f=f_0+\epsilon^2\,f_2\,;\qquad g=\epsilon\,g_1+\epsilon^3\,g_3\,,
\ee
where,
\be \label{eq36}
f_2=\left(A_1\,e^{\,\theta_1+\theta_2}+A_2\, e^{\,\theta_1+\theta_3}+A_3\, e^{\,\theta_2+\theta_3}\right)\,,
\ee
and
\be \label{eq37}
g_1=a\,e^{\,\theta_1}+b\, e^{\,\theta_2}+c\, e^{\,\theta_3}\,; \qquad g_3=C\,e^{\theta_1+\theta_2+\theta_3}\,.
\ee
Here, $a$, $b$, $c$, $A_i, i=1,2,3$ and $C$ are arbitrary constants, $\theta_i=k_i\,x+\omega_i\,t$ for $i=1,2,3$, $f_1=0$, $g_0=0$,  $f_3=0$ and $f_i=g_i=0$ for $i>3$. As in the last case, all the bilinear equations are satisfied when we apply these forms of $f(x,t)$ and $g(x,t)$, for $\omega_i=\pm k_1, i=1,2,3$. 
Thus, one finds three-Galileon solutions that can be written as
\be \label{eq38}
\phi_{\,\pm}=\arctan\,\left[\frac{a\,e^{\theta_1}+b\,e^{\,\theta_2}+c\, e^{\,\theta_3}+C\,e^{\theta_1+\theta_2+\theta_3}}{f_0\!+\!A_1\,e^{\,\theta_1+\theta_2}\!+\!A_2\, e^{\,\theta_1+\theta_3}\!+\!A_3\, e^{\,\theta_2+\theta_3}}\right].
\ee
Again, it is straightforward to check that we can obtain the two-Galileon solution when we set $e^{\,\theta_3}\rightarrow 0$. Now, if we consider $e^{\,\theta_3}\rightarrow \infty$ in the above expression, we obtain
\be
\phi_\pm =\arctan\left[\frac{c+ Ce^{\theta_1+\theta_2}}{A_2 e^{\theta_1}+ A_3 e^{\theta_2}}\right],
\ee
which can be written as in Eq.~(\ref{eq30}), after changing $\theta_1\to-\theta_1$ and making simple manipulations.
For completeness, we illustrate the result (\ref{eq38}) with an example of a three-Galileon, which we depict in Fig. \ref{fig3}.

{\it Ending comments.} In this work we studied the presence of traveling waves in the two-dimensional Galileon theory. The analytical study corroborates the numerical investigations of Ref.~\cite{gs2}. The fact that we can explicitly construct the bilinear equations
\eqref{eq22}-\eqref{eq25}, the symmetry transformations \eqref{eq26a}-\eqref{eq28a}, and two- and three-Galileon solutions strongly suggests that the Galileon traveling waves are solitons, and that the two-dimensional Galileon theory is integrable \cite{HB,H}.
An important physical feature of the Galileon field is to provide calculations in the absence of ghosts, so integrability seems to add nicely, offering another route to further probe the Galileon field value. 

\acknowledgements{The authors would like to thank CAPES and CNPq for partial financial support. They also thank Ashok Das, for driving their attention to the Hirota procedure.}


\begin{thebibliography}{99}
\bb{g1}A. Nicolis, R. Rattazzi, and E. Trincherini, Phys. Rev.  D {\bf 79}, 064036 (2009).
\bb{g2}C. Deffayet, G. Esposito-Farese, and A. Vikman, Phys. Rev.  D {\bf79}, 084003 (2009).
\bb{r1}M. Trodden and K. Hinterbicher, Class. Quan. Grav. {\bf28}, 204003 (2011).
\bb{r2}C. de Ram, Comptes Rendus Physique {\bf 13}, 666 (2012).
\bb{r3}C. Deffayet, D.A. Steer, Class. Quant. Grav. {\bf30}, 214006 (2013).
\bb{gs1}S. Endlich, K. Hinterbichler, L. Hui, A. Nicolis, and J. Wang, JHEP {\bf1105}, 073 (2011).
\bb{gs2}A. Masoumi and Xiao Xiao, Phys. Lett. B {\bf715}, 214 (2012).
\bb{H1}R. Hirota, Phys. Rev. Lett. {\bf27}, 1192 (1971).
\bb{H2}R. Hirota, J. Math Phys. {\bf14}, 810 (1973); J. Hietarinta,
J. Math. Phys. {\bf28}, 1732 (1987); {\bf28}, 2094 (1987); {\bf28}, 2586 (1987); {\bf29}, 628 (1988).
\bb{HI}R. Hirota, Prog. Theo. Phys. {\bf 52}, 1498 (1974).
\bb{HB} R. Hirota, {\it The Direct Method in Soliton Theory} (Cambridge UP, Cambridge, UK, 2004).
\bb{H}J. Hietarinta, Lecture Notes in Physics {\bf767}, 279 (2009).
\bibitem{hz}J. Hietarinta and Da-jun Zhang, Journal of Difference Equations and Applications,
{\bf 19}, 1292 (2013).

\end{thebibliography}
\end{document}